\documentclass[preprint,11pt]{elsarticle}

\usepackage{graphicx}
\usepackage{hyperref}           
\usepackage[english]{babel}
\usepackage{lineno}

\usepackage[utf8]{inputenc} 
\usepackage{siunitx}

\newcommand{\x}{\mathbf{x}}
\renewcommand{\r}{\mathbf{r}}
\newcommand{\grad}{\nabla}

\graphicspath{{figures/}{./}}

\journal{Computer Physics Communications}

\begin{document}

\begin{frontmatter}
\title{How to Differentiate Collective Variables in Free Energy Codes: Computer-Algebra Code Generation and Automatic Differentiation}

\author{Toni Giorgino\fnref{now}}
\fntext[now]{Present address: Istituto di Biofisica, Consiglio Nazionale delle Ricerche (CNR-IBF), 
  c/o Dipartimento di Bioscienze, Università degli Studi di Milano, via Celoria 26, I-20133 Milan, Italy}
\ead{toni.giorgino@cnr.it}
\address{Institute of Neurosciences, National Research Council (CNR-IN),\\ 
  Corso Stati Uniti 4, I-35127, Padua, Italy}


\begin{abstract}
The proper choice of collective variables (CVs) is central to   biased-sampling free energy reconstruction methods in molecular   dynamics simulations. The PLUMED 2 library, for instance, provides  several sophisticated CV choices, implemented in a C++ framework; however, developing new CVs is  still time consuming due to the need to provide code for the analytical   derivatives of all functions with respect to atomic coordinates. We  present two solutions to this problem, namely (a) symbolic   differentiation and code generation, and (b) automatic code   differentiation, in both cases leveraging  open-source   libraries (SymPy and Stan Math respectively). The two approaches are    demonstrated and discussed in detail implementing a realistic   example CV, the local radius of curvature of a polymer. Users may   use the code as a template to streamline the implementation of their   own CVs using high-level constructs and automatic gradient   computation. 
\end{abstract}

\end{frontmatter}
\newpage \noindent
{\bf PROGRAM SUMMARY}

\begin{small}
\noindent
{\em Program Title:}                                          \\
Practical approaches to the differentiation of collective variables in free energy codes: computer-algebra code generation and automatic differentiation \\
{\em Licensing provisions:} \\
GNU Lesser General Public License Version 3 (LGPL-3)                                    \\
{\em Programming languages:} \\
C++, Python                                   \\
{\em Nature of problem:}\\
The C++ implementation of collective variables (CVs, functions of
atomic coordinates to be used in biased sampling applications) in
biasing libraries for atomistic simulations, such as PLUMED [1],
requires computation of both the variable and its gradient with
respect to the atomic coordinates; coding and testing the analytical
derivatives complicates  the implementation of new CVs.\\
{\em Solution method:}\\
The paper shows two approaches to automate the computation of CV
gradients, namely, symbolic differentiation with code generation and
automatic code differentiation, demonstrating their implementation
entirely with open-source software (respectively, SymPy and the Stan Math Library). \\
{\em Additional comments:}\\
The paper's accompanying code serves as an example and template for
the methods described in the paper; it is distributed as the two
modules \texttt{curvature\_codegen} and \texttt{curvature\_autodiff}
integrated in PLUMED 2's source tree; the latest version is available
at \url{https://github.com/tonigi/plumed2-automatic-gradients} .
\\

\end{small}


\section{Introduction}

Biased approaches to molecular dynamics (MD) enable the sampling of
events whose occurrence would otherwise be prohibitively rare on the
time scales affordable by direct (unbiased) integration of the
equations of motion. Central to the possibility to obtain converged
estimates of thermodynamic quantities is the search of appropriate
projections of the system state~\cite{Laio_Gervasio_2008}; in turn,
this enables the search of a reaction coordinate to effectively push
the specific system out of free energy minima.  When an appropriate
reaction coordinate is selected, the sampling of a system can be
accelerated through a number of \emph{biased} sampling methods, such
as umbrella sampling~\cite{torrie_nonphysical_1977},
metadynamics~\cite{laio_escaping_2002},
SuMD~\cite{salmaso_exploring_2017} and
others~\cite{hamelberg_accelerated_2004}, most of which enable the
reconstruction of the free energy landscape, and in some cases the
kinetics~\cite{mollica_kinetics_2015,sun_characterizing_2017}, on the
space spanned by the chosen variables.  The availability of a wide
range of functions of atomic coordinates (collective variables or CVs)
is thus a valuable asset in the construction of proper reaction
coordinates~\cite{giorgino_metagui_2017}.

\begin{figure}
  \centering
  \includegraphics[width=.8\textwidth]{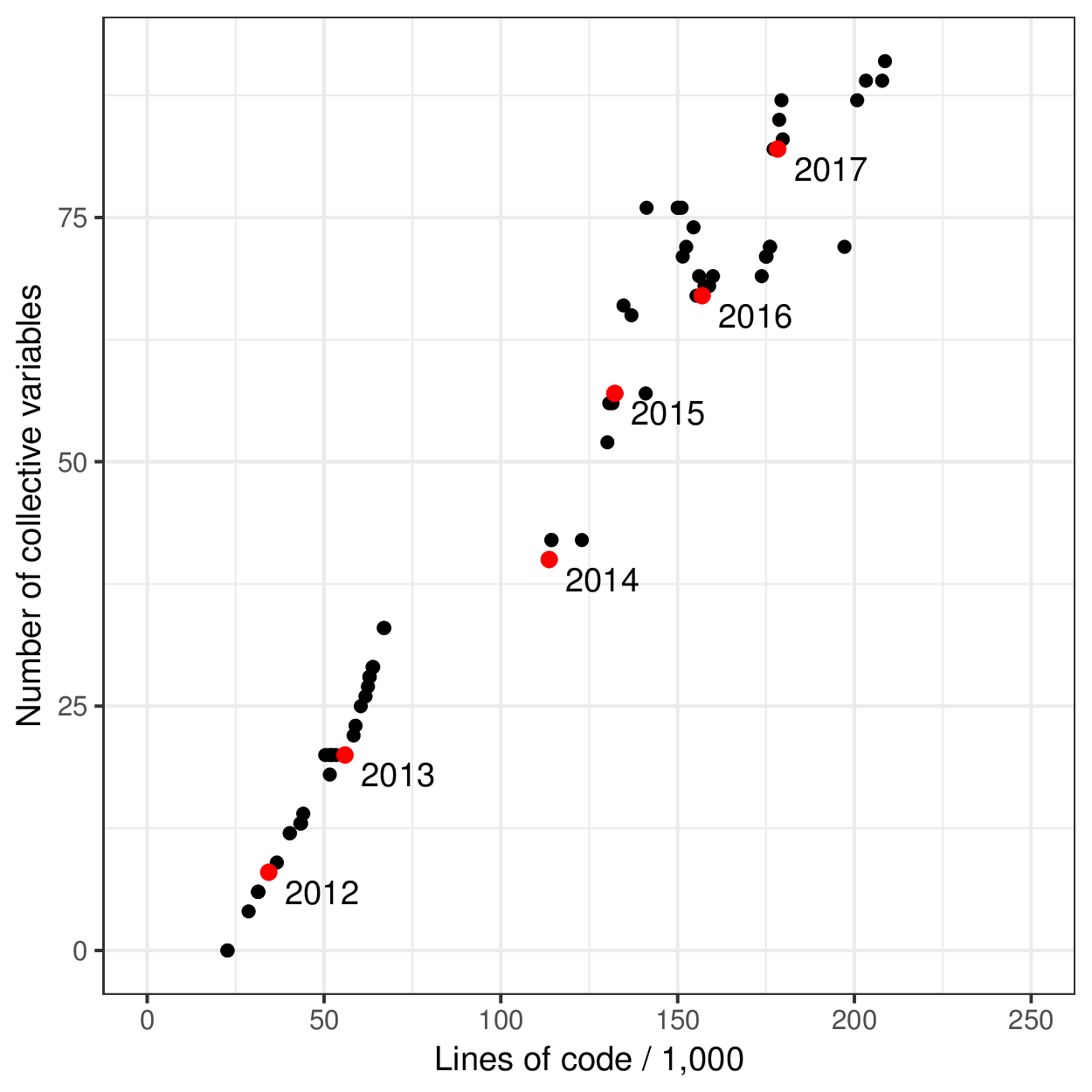}
  \caption{Growth of the number of CVs in PLUMED~2 and
    the corresponding code base (lines of C++ code in the master
    branch, including headers and inline documentation; the 
    count also includes support functions, command 
    line utilities and file readers). }
  \label{fig:counts}
\end{figure}

PLUMED~2 is a widely-used engine to perform biased sampling
simulations in atomistic simulations~\cite{Tribello_Bonomi_Branduardi_Camilloni_Bussi_2014}.
Part of PLUMED's success is due to the number and variety of
collective variables implemented (see
e.g.~\cite{tribello_analyzing_2017,branduardi_b_2007,bonomi_integrative_2017}),
enabling projections of the system state on ``axes'' of intuitive
value, and the number of CVs implemented in PLUMED~2 has been growing
steadily (Figure~\ref{fig:counts}).  Users can incorporate their own
CVs coding them in C++; however, the implementation of complex
functions is complicated by the need to compute gradients with respect
to atomic coordinates, which increases the complexity and debugging
time of the corresponding source codes.

Here, we present two approaches to automatically implement
CV gradients:
\begin{enumerate}
\item a \emph{symbolic differentiation with code generation}
  approach, where the SymPy computer algebra system (CAS)~\cite{meurer_sympy:_2017} is used to
  derive the expressions and automatically produce an equivalent C
  function (Section~\ref{sec:generating-code-with}); and
\item an \emph{automatic code differentiation} approach, using the
  reverse-mode code differentiation capabilities provided by the Stan
  Math library~\cite{carpenter_stan_2015} (Section~\ref{sec:autom-code-diff}).
\end{enumerate}

We demonstrate the two approaches on the simple (yet non trivial) case
of a CV computing the local curvature of a polymer, approximated as
the radius of curvature of a circle passing through three consecutive
atoms. The two approaches provide identical numerical
results and are based on mature and well-known open
source libraries. Their  different characteristics will be presented in the
discussion section. 

Example code is made available as
open-source respectively in PLUMED~2's \texttt{curvature\_codegen} and
\texttt{curvature\_autodiff} modules; from there, the corresponding source files
can be used as templates for the implementation of customized CVs.

\subsection{Background}

A CV is a function of a system's state through the coordinates of its
$n$ particles, namely:~\cite{fiorin_using_2013}
\[
 f(\x) = f(\x_1, \ldots, \x_n)
\]

\noindent Applying biases to CVs implies that the system is subject to a
potential $V$ which depends on the coordinates solely through $f$: 
\[
 V(\x) = V_1(f(\x))
\]

The bias potential translates to forces acting on each atom, which are
computed in the biasing library and passed to the molecular dynamics (MD)
engine. The MD code adds them to those due to the force field, and
integrates the equations of motion. From the chain differentiation
rule,
\[
\mathbf{F}(\x) = -\grad_\x V_1(f(\x)) = 
     - \frac{\partial V_1(f)}{\partial f} \grad_\x f(\x)
\]
here $\partial V_1 / \partial f$ is the \emph{generalized force}, 
set by the chosen biasing scheme (e.g., a time-dependent sum of
Gaussians in the case of metadynamics), while  $\grad_\x f$ 
depends only on the functional form of $f$ and the
system state $\x$.
Implementation of a CV requires the programmer to write code
for  $f(\x)$ and its derivatives with respect to all of the
arguments (number of involved atoms times three Cartesian components).

\subsection{Radius of curvature}

\begin{figure}
  \centering
  \includegraphics[width=.5\textwidth]{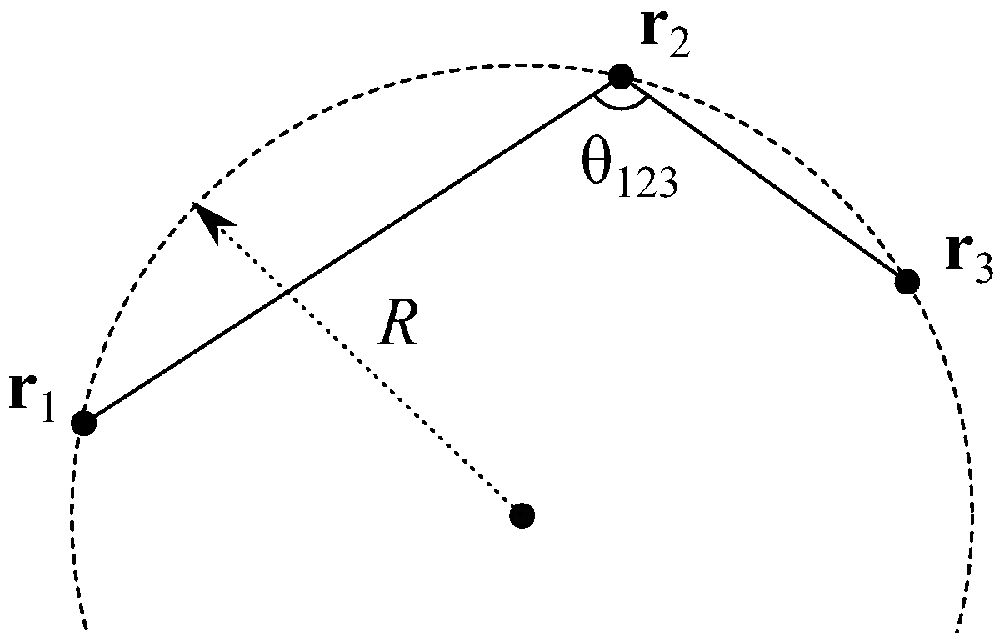}
  \caption{The radius of curvature  as a collective
    variable $R(\r_1,\r_2,\r_3)$.}
  \label{fig:radius}
\end{figure}

To illustrate the methods, we shall use as an example
the radius of curvature at a given atom along a
polymer.  A natural choice for this quantity is to compute the radius
$R$ of the circle (circumcircle) passing through three given points
$\r_1, \r_2$ and $\r_3$ (Figure~\ref{fig:radius}), e.g.\ the centers of consecutive
C$\alpha$ atoms. The diameter $2R$
is obtained elementarily  via the sine rule as  the ratio between
a side of the triangle formed by the  points and the sine of the opposing angle, i.e.,
calling $\r_{ij} = \r_i-\r_j$ and $\theta_{123}$ the angle at $\r_2$,
\begin{equation}\label{eq:R}
 2R  = \frac { |\r_{13}| }{\sin \theta_{123}} \qquad \mbox{with} \qquad
 \cos \theta_{123} = \frac{ \r_{12} \cdot \r_{23} }{  |\r_{12} | |\r_{23} | }
\end{equation}

The above expression is compact  in vector notation, but
the expressions for its gradient in
Cartesian coordinates, i.e. the components of
$\grad_\x R(\r_1, \r_2, \r_3)$ with
$\x=(r_{1x},r_{1y}, \ldots, r_{3z})$, are unwieldy (see 
the notebook \texttt{CurvatureCodegen.ipynb}).

\subsection{Edge cases and inverse radius}

Computer-assisted code generation does not automatically guarantee
that the functions are well-defined in all conditions. Of special
relevance are singularities on edge cases, such as
collinearity ($R \to \infty$) in the curvature example.  Edge cases are generally set
aside when deriving expressions ``on paper'', but their occurrence in
computer code, however rare, must be caught  to avoid
crashes in  simulations.  


In the example of this paper  the user-visible
\texttt{INVERSE} flag is added to the curvature collective
variables in order to illustrate a possible approach to removing 
singularities, and to show how CV computations can
be made to depend on user-defined parameter. The idea is that the reciprocal of the radius is a
better-behaved collective variable, lacking the singularity (infinite
radius) for the case of three collinear points (which may arise
e.g. when the initial configuration of a polymer is generated
artificially). Of note, this solution does not eliminate a 
singularity in the gradient, whose limit  for 
collinear atoms is still undefined.

\section{Generating code with a computer algebra system}\label{sec:generating-code-with}

\begin{figure}
  \centering
  \includegraphics[width=.9\textwidth]{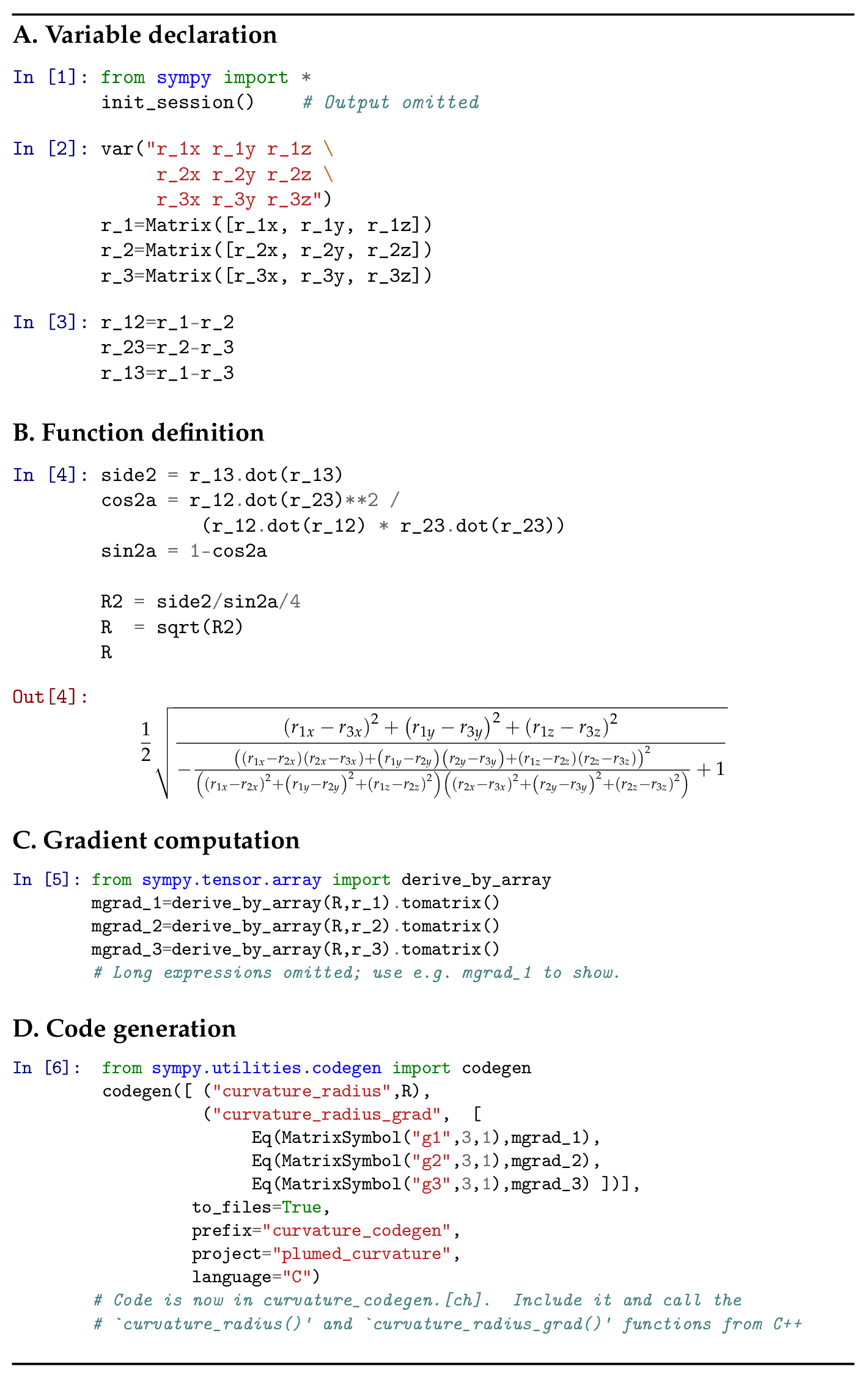}
  \caption{\emph{Symbolic differentiation with code generation
      approach} -- A SymPy Jupyter notebook generating C code for the
    collective variable \texttt{R} and its gradient.  (See also the
    calling code in the module \texttt{curvature\_codegen/Curvature.cpp}, and
    the notebook \texttt{CurvatureCodegen.ipynb} containing the
    extended version of this figure.)  }
  \label{fig:codegeneration}
\end{figure}

CAS manipulate mathematical expressions in
 symbolic form. They usually build internal representations of
expression as trees, which are subject to transformations
encoding algebraic manipulations (such as differentiation)
as pattern-matching rules.  When desired, the trees can
be evaluated numerically, printed, or transformed in other languages.
It is therefore tempting to use CAS to generate lengthy mathematical
expressions for later compilation and inclusion in CV code.

We used the SymPy package~\cite{meurer_sympy:_2017} to  implement
the collective variable as a mathematical expression,  compute its
gradient symbolically, and  convert the function and the gradient
into C code for inclusion in PLUMED.  
SymPy turned out to be suitable for this task because of
three reasons: it is open source and freely available; it provides
excellent symbolic manipulation and simplification features (among
others); and generates stand-alone C code which does not rely on
external libraries. It has been used for code-generation purposes
in other contexts~\cite{mushtaq_automatic_2014}.

\renewcommand{\theenumi}{\Alph{enumi}} 

Figure \ref{fig:codegeneration} shows the steps required for this
approach, i.e.: 
\begin{enumerate}
\item Atom coordinates are introduced as SymPy
symbols. 
\item The collective variable is defined following
eq.~(\ref{eq:R}); note the use of vector algebra.  
\item Symbolic
computation of $\grad_\x R(\r_1,\r_2,\r_3)$. 
\item Code generation is
performed with the \texttt{codegen()} function, which translates
\texttt{R} and \texttt{mgrad} in the files
\texttt{curvature\_codegen.[ch]}. 
\end{enumerate}

The code generated is included via a wrapper, which makes the
functions \texttt{curvature\_radius} and
\texttt{curvature\_radius\_grad} available for use in the
\texttt{Curvature} class.  The rest of the code does not depend on the
specific CV function, and can be reused from the example's source
(\texttt{Curvature.cpp)},  available in PLUMED's
\texttt{curvature\_codegen} module. The module also contains a
\emph{multicolvar} implementation, enabling the use of aggregated
curvature radius values along a polymer (e.g., its mean, minimum and
so on).

An extended version of the notebook of Figure~\ref{fig:codegeneration}
distributed with the source code also demonstrates how substitution
operators were used within the CAS to check the results of the
derivations with respect to known values and limits towards edge cases
(in this case, collinear points).  There was
no need to generate separate expressions for the inverse radius, whose
gradient is trivially implemented in C++ via the chain rule.

Finally, depending on the symmetry of the CV and the number of atoms
involved, it may be more natural to differentiate with respect to
atoms' distance vectors rather than coordinates, and then apply the
chain rule in the calling code.  The code generation steps proceed
straightforwardly as above (final examples in the notebook).

\subsection{Symbolic Common Subexpression Elimination}\label{sec:symb-comm-subexpr}

Inspection of the code generated by the ``naive'' CAS
approach in Figure~\ref{fig:codegeneration} shows repeatedly-computed
expressions that could be made more efficient with the introduction of
intermediate variables.  This is due to the fact that the translation
of a formula derived by contemporary CAS systems into code form
usually happens all at once on the basis of the explicit expression,
which may be unnecessarily (or even intractably) complex; in
particular, the translation does not re-use the subexpressions which are generated
during differentiation. As an example, consider the derivative
$\frac{d}{dx} \exp(\exp(f(x)) = \exp(\exp(f(x))) \exp(f(x)) f'(x) $:
even though the $\exp(f(x))$ subexpression could in principle be
evaluated just once, this simplification can not be rendered in the
mathematical formula.

SymPy can indeed exploit opportunities for \emph{common subexpression
  elimination} (CSE) at a symbolic level, as demonstrated in the
section \emph{Common subexpression elimination} of the notebook.  The
source code of the gradient function generated by \verb+codegen()+
contains approximately 1,980 floating-point (FP) operators.  The
symbolic CSE step, in contrast, produces the source code of an
equivalent gradient function containing just 101 FP operators (the
radius function has 44).

It is important to note that most of the redundant arithmetic operations are
optimized by the compiler anyway, because CSE is a standard pass in current
compiler optimizations; however, the extent of subexpressions that
are going to be recognized at this low-level pass is hard to assess
\emph{a priori}. In Section~\ref{sec:perf-cons} we report actual FP counts
measured on naive and CSE code downstream  of the compiler optimization
passes.

\section{Automatic code differentiation}\label{sec:autom-code-diff}

\begin{figure}
  \centering
  \includegraphics[width=.8\textwidth]{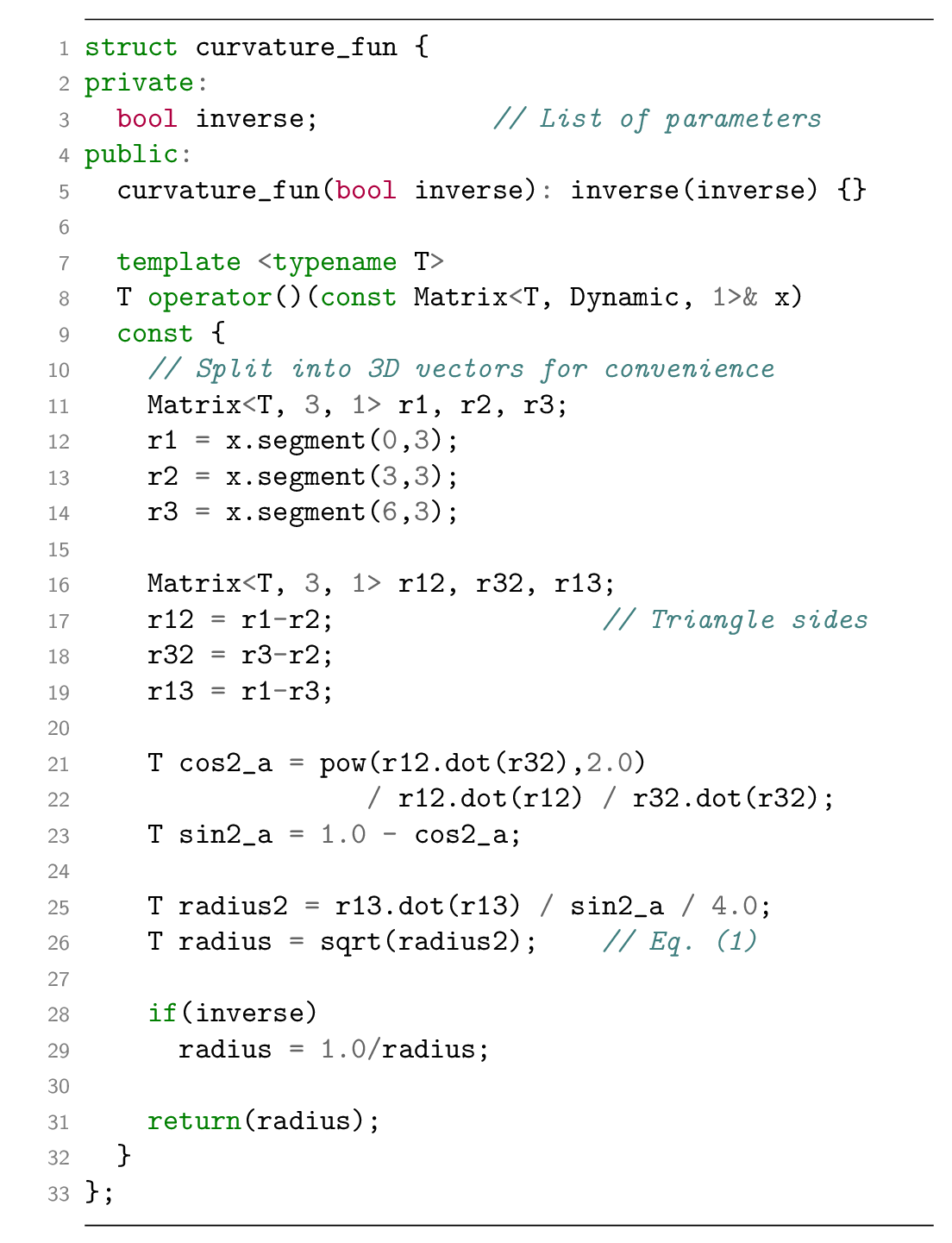}
  \caption{\emph{Automatic code differentiation approach} -- Automatically
    differentiable code implementing the radius of curvature CV.  The
    ``abstract'' type \texttt{T} is used for variables and
    parameters. The use of \texttt{Eigen} types allows writing
    Eq.~(\ref{eq:R}) in a compact vector form close to the textbook
    one. See \texttt{CurvatureAutoDiff.cpp} in the PLUMED module
    \texttt{curvature\_autodiff} for full code.}
  \label{fig:functor}
\end{figure}

A different and independent approach to gradient computation is through
\emph{automatic code differentiation}, a powerful method which calculates 
gradients of functions defined by (in principle) arbitrary
algorithms.  In short, the gradient computation ``mirrors'' each
elementary step executed by the function being derived by keeping
track of the partial derivatives (``adjunct'') of each variable,
propagating them via the chain rule; the components of a gradient are
computed together in a single pass. It has the same computational
complexity as the original code and hence, for example, loops of
arbitrary length can be differentiated even when the
number of iterations is only known at run time
(see~\cite{carpenter_stan_2015} for a thorough
explanation).

We rely on Stan Math, a header-only library part of the Stan
probabilistic programming language, to provide reverse-mode automatic
differentiation for C++ code~\cite{stanJSS,carpenter_stan_2015}.  The
library uses template-based
metaprogramming, meaning that expressions can be written with the same
semantics used for common floating-point operations, while in reality
operator overloading is used to construct the code and data structures
necessary for differentiation.

In practice, a convenient way to introduce this in PLUMED is to wrap
the function to be differentiated in a C++ functor, as shown in
Figure~\ref{fig:functor}.  Apart from  boilerplate
semantics, the body of function can be written as customary, with the
following assumptions:
\begin{itemize}
\item arguments are passed as one single-column \texttt{Matrix}
  object (typically as long as the number of atomic coordinates
  involved in the collective variable);
\item arguments and return value should all be of ``type'' \texttt{T};
\item parameters, if needed, can passed through the functor constructor
  (see section~\ref{sec:deal-with-param} for details).
\end{itemize}

The gradient computation is transparent for the programmer. The
collective variable's \texttt{compute()} method, invoked by PLUMED at
each time step of the simulation, will call
\texttt{stan::math::gradient()}, perform any vector reshaping
necessary, then set the derivatives to inform the rest of the code of
the bias forces to be applied. The \texttt{compute} method 
is thus  largely
independent on the CV at hand, and can be re-used unchanged from the
complete example provided in the \texttt{CurvatureAutoDiff.cpp} file.

\subsection{Linear algebra and special functions}

The example code  expresses the curvature function concisely by using 
\texttt{dot()} inner product operators; they operate
 between two $3 \times 1$
vectors (the sides of the triangle), declared of type
\texttt{Eigen::Matrix<T,1,3>}.  Eigen, a template-based linear algebra
library, defines operations on dense and sparse vectors and matrices,
and is part of Stan Math.  Its scope is in fact  much larger, providing
e.g.\ determinants, eigenvectors and many decomposition types~\cite{eigenweb}.

In addition to the code differentiation features,  the Stan Math
library provides  a remarkable range of special functions and
distributions, whose enumeration is beyond the scope of this paper.
Functions can also be defined through differential equations and
differentiated with respect to their parameters, thanks to SUNDIALS'
CVodes library~\cite{carpenter_stan_2015,stanJSS,hindmarsh_sundials:_2005}.

\subsection{Dealing with parameters} \label{sec:deal-with-param}


Functors expect only an N-dimensional vector of variables; lines 3--5
of Figure~\ref{fig:functor} show how additional parameters (e.g. set
by the user in PLUMED's input file) can be passed to the functor
constructor.

Note that parameters should not be passed via the arguments (e.g., as
variables in addition to the atom coordinates) because, besides
restricting their types to floating point numbers, the function
would be unnecessarily differentiated with respect to them (although
reverse mode differentiation is relatively efficient in this regard).

\section{Performance considerations}\label{sec:perf-cons}

In order to estimate real-world performance, i.e.\ downstream of
optimizations performed automatically by modern compilers, we compared
the optimized executables
in terms of FP instructions actually executed by the CPU. Measurements
were obtained through hardware FP counters, namely
\texttt{FP\_COMP\_OPS\_EXE:SSE\_SCALAR\_DOUBLE} and  vector
analogs.  Table~\ref{tab:perf}
reports the performance of the functions performing the radius and
gradient computations, in terms of number of double-precision FP
instructions, and CPU time required per calculation (medians and
inter-quartile ranges over 10 runs of $10^7$ evaluations each). 

In Section~\ref{sec:symb-comm-subexpr} we reported that the number of
FP operators found in the naive and symbolic CSE-generated
\emph{source} codes differed by a factor of 20.  The dramatic
difference is not reflected in the \emph{compiled} code, indicating
that the compiler optimized out almost all of the redundant
expressions.  The high-level CSE step was still beneficial, reducing
the final FP instruction count by further 45 (out of 137).

Interestingly, the automatic differentiation approach used roughly the
same number of floating-point operations as the CAS-generated
code. Its actual run time was longer, likely due to non-FP
operations such as method calls, expected because of the use of
objects representing variables along with their adjoints. Note also that
run times varied widely with ``fast-math''-like optimizations, and
that hardware counters may be affected by
approximations~\cite{weaver_non-determinism_2013}.

Finally, we remark that all of the above measurements included the
FP-intensive part of the calculations only. If speed is a concern, the
overhead at the interface with the free-energy code (e.g., reshaping
coordinate arrays) should be accounted, as it may dwarf the cost of
the functions proper.


\begin{table}
  \centerline{
    \sisetup{separate-uncertainty}
  \begin{tabular}{lccSS}
    \hline
                                  & \multicolumn{2}{c}{Floating-point ops.} & \multicolumn{2}{c}{CPU time (ns)}                         \\
    Method                        & {$R$ and $\grad R$}                     & {$\grad R$ only} & {$R$ and $\grad R$} & {$\grad R$ only} \\
    \hline
    Code generation               & 173                                     & 137              & 125.98+- 0.07       & 110.36 +- 0.02    \\
    Code generation, symbolic CSE & 128                                     & 92               & 60.19 +- 0.01       & 48.48 +- 0.11    \\
    Automatic differentiation     & 148                                     & {---}            & 532.61+- 7.82       & {---}            \\
    \hline
  \end{tabular}
  }
  \caption{Number of floating-point operations and CPU time required
    by one evaluation of the gradient of the curvature function with
    the approaches presented in the paper. One iteration of either the
    curvature function and its gradient, or gradient only, as
    indicated in the column headers, is measured; time spent in the
    interface with the free energy code is not included. (Medians and
    inter-quartile ranges over $10 \times 10^7$ calls; Intel Xeon CPU
    E5-2697 v4 at 2.30GHz, GCC version 6.1.0, optimization O3.) }
  \label{tab:perf}
\end{table}


\section{Discussion}

The two approaches discussed yield, as expected, numerical results
equal within machine precision. Which of the two is preferred 
depends on the complexity of the specific problem being addressed.

On the one hand, symbolic differentiation may be closer to the
``classroom'' approach, where a closed form expression is derived
top-down. CAS assist  the
development of the final formulas, e.g. enabling complex
substitutions, differentiations, and simplifications. Symbols can also
be replaced by values and evaluated at any time necessary, which is
generally useful for quickly checking the consistency of equations
with known results. Also, structuring computations as ``notebook'' format,
while optional, is an approach favoured by most CAS
as a means to keep a readable and reproducible
record of the steps leading to a particular formula or
code~\cite{perez_ipython:_2007}. 

The most important limitation of CAS code generators is that they do
not handle generic functions defined as algorithms (e.g., loops until
convergence). As discussed above, the automatic code differentiation
approach largely solves the issue: it is therefore expected to be the
preferred way to implement very complex CVs in real-world
problems. The vastly increased generality comes at some expense of
convenience, for the edit-compile-run cycle is somewhat at odds with
notebook-style readability and the interactive testing it affords.
Performance-wise, the approaches are very close in terms of amount of
the floating-point calculations required; and at least in the same
order of magnitude in terms of CPU time.

An even more high-level language approach than the ones presented here
would be to evaluate mathematical functions in an embedded Python
interpreter, an  approach recently implemented in PLUMED (also
used in~\cite{galvelis_neural_2017}). This may be desirable
for casual coding, but likely inefficient, as the critical portion of
the calculations would happen in an interpreted language.

The code templates presented have been tested in combination with
software version widely in use at the time of writing
(Table~\ref{tab:versions}). The
libraries are under active development, so  code may require
minor adaptations with future releases.  Regression testing and
continuous integration of the PLUMED code base ensure that
incompatibilities, should they arise, will be spotted timely.

\begin{table}
  \centering
  \begin{tabular}{cc}
    \hline
    Software          & Version \\
    \hline
    Plumed            & 2.4.0   \\
    Python (Anaconda) & 3.6.4   \\
    SymPy             & 1.0     \\
    Stan Math library & 2.16.0  \\
    GCC               & 6.1.0   \\
    Clang++           & 3.8.1   \\
    Intel ICC         & 18.0.1  \\ 
    \hline
  \end{tabular}
  \caption{Software versions tested.}
  \label{tab:versions}
\end{table}

\section{Conclusion}

This paper presented two methods intended to substantially reduce the
barrier to the development of functions of atomic coordinates.  While
the resulting code may not as optimized as hand-written one, it is
hoped that the approaches presented will enable the extension of free
energy codes with further CVs, significant from the points of view of
structural biology, biological relevance, or closeness to experimental
observables, whose complexity would have otherwise made their
implementation prohibitive.

\section{Acknowledgements}

I would like to thank Prof.\ G.\ Bussi and Prof.\ C.\ Camilloni for
discussions on the applications of automatic differentiation and
comments on the manuscript.  I acknowledge CINECA awards under the
ISCRA initiative for the availability of high performance computing
resources and support. Research funding from Acellera Ltd.\ is
gratefully acknowledged.

%

\section{References}

\bibliographystyle{elsarticle-num}

\begin{thebibliography}{0}
\bibitem{1} Tribello GA, Bonomi M, Branduardi D, Camilloni C, Bussi
  G. PLUMED~2: New feathers for an old bird. Computer Physics
  Communications. 2014 Feb;185(2):604-13.

\end{thebibliography}

\begin{thebibliography}{10}
\expandafter\ifx\csname url\endcsname\relax
  \def\url#1{\texttt{#1}}\fi
\expandafter\ifx\csname urlprefix\endcsname\relax\def\urlprefix{URL }\fi
\expandafter\ifx\csname href\endcsname\relax
  \def\href#1#2{#2} \def\path#1{#1}\fi

\bibitem{Laio_Gervasio_2008}
A.~Laio, F.~L. Gervasio, Metadynamics: a method to simulate rare events and
  reconstruct the free energy in biophysics, chemistry and material science,
  Reports on Progress in Physics 71~(12) (2008) 126601.
\newblock \href {http://dx.doi.org/10.1088/0034-4885/71/12/126601}
  {\path{doi:10.1088/0034-4885/71/12/126601}}.

\bibitem{torrie_nonphysical_1977}
G.~M. Torrie, J.~P. Valleau, Nonphysical sampling distributions in {Monte}
  {Carlo} free-energy estimation: {Umbrella} sampling, Journal of Computational
  Physics 23~(2) (1977) 187--199.
\newblock \href {http://dx.doi.org/10.1016/0021-9991(77)90121-8}
  {\path{doi:10.1016/0021-9991(77)90121-8}}.

\bibitem{laio_escaping_2002}
A.~Laio, M.~Parrinello, Escaping free-energy minima, Proceedings of the
  National Academy of Sciences of the United States of America 99~(20) (2002)
  12562--12566.
\newblock \href {http://dx.doi.org/10.1073/pnas.202427399}
  {\path{doi:10.1073/pnas.202427399}}.

\bibitem{salmaso_exploring_2017}
V.~Salmaso, M.~Sturlese, A.~Cuzzolin, S.~Moro, Exploring {Protein}-{Peptide}
  {Recognition} {Pathways} {Using} a {Supervised} {Molecular} {Dynamics}
  {Approach}, Structure 25~(4) (2017) 655--662.e2.
\newblock \href {http://dx.doi.org/10.1016/j.str.2017.02.009}
  {\path{doi:10.1016/j.str.2017.02.009}}.

\bibitem{hamelberg_accelerated_2004}
D.~Hamelberg, J.~Mongan, J.~A. McCammon, Accelerated molecular dynamics: {A}
  promising and efficient simulation method for biomolecules, The Journal of
  Chemical Physics 120~(24) (2004) 11919--11929.
\newblock \href {http://dx.doi.org/10.1063/1.1755656}
  {\path{doi:10.1063/1.1755656}}.

\bibitem{mollica_kinetics_2015}
L.~Mollica, S.~Decherchi, S.~R. Zia, R.~Gaspari, A.~Cavalli, W.~Rocchia,
  Kinetics of protein-ligand unbinding via smoothed potential molecular
  dynamics simulations, Scientific Reports 5 (2015) srep11539.
\newblock \href {http://dx.doi.org/10.1038/srep11539}
  {\path{doi:10.1038/srep11539}}.

\bibitem{sun_characterizing_2017}
H.~Sun, Y.~Li, M.~Shen, D.~Li, Y.~Kang, T.~Hou, Characterizing
  {Drug}–{Target} {Residence} {Time} with {Metadynamics}: {How} {To}
  {Achieve} {Dissociation} {Rate} {Efficiently} without {Losing} {Accuracy}
  against {Time}-{Consuming} {Approaches}, Journal of Chemical Information and
  Modeling\href {http://dx.doi.org/10.1021/acs.jcim.7b00075}
  {\path{doi:10.1021/acs.jcim.7b00075}}.

\bibitem{giorgino_metagui_2017}
T.~Giorgino, A.~Laio, A.~Rodriguez, {METAGUI} 3: {A} graphical user interface
  for choosing the collective variables in molecular dynamics simulations,
  Computer Physics Communications 217 (2017) 204--209.
\newblock \href {http://dx.doi.org/10.1016/j.cpc.2017.04.009}
  {\path{doi:10.1016/j.cpc.2017.04.009}}.

\bibitem{Tribello_Bonomi_Branduardi_Camilloni_Bussi_2014}
G.~A. Tribello, M.~Bonomi, D.~Branduardi, C.~Camilloni, G.~Bussi, Plumed 2: New
  feathers for an old bird, Computer Physics Communications 185~(2) (2014)
  604–613.
\newblock \href {http://dx.doi.org/10.1016/j.cpc.2013.09.018}
  {\path{doi:10.1016/j.cpc.2013.09.018}}.

\bibitem{tribello_analyzing_2017}
G.~A. Tribello, F.~Giberti, G.~C. Sosso, M.~Salvalaglio, M.~Parrinello,
  Analyzing and {Driving} {Cluster} {Formation} in {Atomistic} {Simulations},
  Journal of Chemical Theory and Computation 13~(3) (2017) 1317--1327.
\newblock \href {http://dx.doi.org/10.1021/acs.jctc.6b01073}
  {\path{doi:10.1021/acs.jctc.6b01073}}.

\bibitem{branduardi_b_2007}
D.~Branduardi, F.~L. Gervasio, M.~Parrinello, From {A} to {B} in free energy
  space, The Journal of Chemical Physics 126~(5) (2007) 054103.
\newblock \href {http://dx.doi.org/10.1063/1.2432340}
  {\path{doi:10.1063/1.2432340}}.

\bibitem{bonomi_integrative_2017}
M.~Bonomi, C.~Camilloni, Integrative structural and dynamical biology with
  {PLUMED}-{ISDB}, Bioinformatics 33~(24) (2017) 3999--4000.
\newblock \href {http://dx.doi.org/10.1093/bioinformatics/btx529}
  {\path{doi:10.1093/bioinformatics/btx529}}.

\bibitem{meurer_sympy:_2017}
A.~Meurer, C.~P. Smith, M.~Paprocki, O.~Čertík, S.~B. Kirpichev, M.~Rocklin,
  A.~Kumar, S.~Ivanov, J.~K. Moore, S.~Singh, T.~Rathnayake, S.~Vig, B.~E.
  Granger, R.~P. Muller, F.~Bonazzi, H.~Gupta, S.~Vats, F.~Johansson,
  F.~Pedregosa, M.~J. Curry, A.~R. Terrel, S.~Roučka, A.~Saboo, I.~Fernando,
  S.~Kulal, R.~Cimrman, A.~Scopatz, {SymPy}: symbolic computing in {Python},
  PeerJ Computer Science 3 (2017) e103.
\newblock \href {http://dx.doi.org/10.7717/peerj-cs.103}
  {\path{doi:10.7717/peerj-cs.103}}.

\bibitem{carpenter_stan_2015}
B.~Carpenter, M.~D. Hoffman, M.~Brubaker, D.~Lee, P.~Li, M.~Betancourt, The
  {Stan} {Math} {Library}: {Reverse}-{Mode} {Automatic} {Differentiation} in
  {C}++, arXiv:1509.07164 [cs]ArXiv: 1509.07164.

\bibitem{fiorin_using_2013}
G.~Fiorin, M.~L. Klein, J.~Hénin, Using collective variables to drive
  molecular dynamics simulations, Molecular Physics 111~(22-23) (2013)
  3345--3362.
\newblock \href {http://dx.doi.org/10.1080/00268976.2013.813594}
  {\path{doi:10.1080/00268976.2013.813594}}.

\bibitem{mushtaq_automatic_2014}
A.~Mushtaq, K.~Olaussen, Automatic code generator for higher order integrators,
  Computer Physics Communications 185~(5) (2014) 1461--1472.
\newblock \href {http://dx.doi.org/10.1016/j.cpc.2014.01.012}
  {\path{doi:10.1016/j.cpc.2014.01.012}}.

\bibitem{stanJSS}
B.~Carpenter, A.~Gelman, M.~Hoffman, D.~Lee, B.~Goodrich, M.~Betancourt,
  M.~Brubaker, J.~Guo, P.~Li, A.~Riddell, Stan: A probabilistic programming
  language, Journal of Statistical Software, Articles 76~(1) (2017) 1--32.
\newblock \href {http://dx.doi.org/10.18637/jss.v076.i01}
  {\path{doi:10.18637/jss.v076.i01}}.

\bibitem{eigenweb}
G.~Guennebaud, B.~Jacob, et~al., Eigen v3, http://eigen.tuxfamily.org (2010).

\bibitem{hindmarsh_sundials:_2005}
A.~C. Hindmarsh, P.~N. Brown, K.~E. Grant, S.~L. Lee, R.~Serban, D.~E.
  Shumaker, C.~S. Woodward, {SUNDIALS}: {Suite} of {Nonlinear} and
  {Differential}/{Algebraic} {Equation} {Solvers}, ACM Trans. Math. Softw.
  31~(3) (2005) 363--396.
\newblock \href {http://dx.doi.org/10.1145/1089014.1089020}
  {\path{doi:10.1145/1089014.1089020}}.

\bibitem{weaver_non-determinism_2013}
V.~M. Weaver, D.~Terpstra, S.~Moore, Non-determinism and overcount on modern
  hardware performance counter implementations, in: 2013 {IEEE} {International}
  {Symposium} on {Performance} {Analysis} of {Systems} and {Software}
  ({ISPASS}), 2013, pp. 215--224.
\newblock \href {http://dx.doi.org/10.1109/ISPASS.2013.6557172}
  {\path{doi:10.1109/ISPASS.2013.6557172}}.

\bibitem{perez_ipython:_2007}
F.~Perez, B.~E. Granger, {IPython}: {A} {System} for {Interactive} {Scientific}
  {Computing}, Computing in Science Engineering 9~(3) (2007) 21--29.
\newblock \href {http://dx.doi.org/10.1109/MCSE.2007.53}
  {\path{doi:10.1109/MCSE.2007.53}}.

\bibitem{galvelis_neural_2017}
R.~Galvelis, Y.~Sugita, Neural {Network} and {Nearest} {Neighbor} {Algorithms}
  for {Enhancing} {Sampling} of {Molecular} {Dynamics}, Journal of Chemical
  Theory and Computation 13~(6) (2017) 2489--2500.
\newblock \href {http://dx.doi.org/10.1021/acs.jctc.7b00188}
  {\path{doi:10.1021/acs.jctc.7b00188}}.

\end{thebibliography}

\end{document}